\documentclass[12pt]{article}

\usepackage[margin=2.54cm]{geometry}
\usepackage{lineno}                        
\usepackage{setspace}

\usepackage[english]{babel}
\usepackage[T1]{fontenc}
\usepackage[]{times}
\usepackage{amsmath,amssymb,amsthm,stmaryrd}
\usepackage{bm,bbm}
\usepackage{natbib}
\usepackage{graphicx}
\usepackage{comment,ulem}
\usepackage{array}
\usepackage{longtable,dcolumn,booktabs,mathtools}
\usepackage{url}
\usepackage{pdflscape}
\usepackage{pdfpages}
\usepackage{paralist}
\usepackage{authblk}
\usepackage{setspace}
\usepackage{textcomp}
\newcolumntype{d}[1]{D{.}{.}{#1}}

\makeatletter
\newcommand\appendix@section[1]{%
  \refstepcounter{section}%
  \orig@section*{Appendix \@Alph\c@section: #1}%
  \addcontentsline{toc}{section}{Appendix \@Alph\c@section: #1}%
}
\let\orig@section\section
\g@addto@macro\appendix{\let\section\appendix@section}
\makeatother

\usepackage{caption}
\newcommand{\real}{\mathbb{R}}

\newcommand{\E}{\mathbb{E}}

\newcommand{\dd}{\textup{d}}
\newcommand{\fx}{\mbox{\sl fx}}
\newcommand{\ff}{\mbox{\sl ff}}
\newcommand{\ffVar}{\mbox{\sl ffVar}}
\newcommand{\rr}{\mbox{\sl rr}}
\newcommand{\rMax}{\mbox{\sl rMax}}
\newcommand{\dP}{\mbox{\sl dP}}
\newcommand{\slP}{\mbox{\sl P}}
\newcommand{\xff}{x_{\mbox{\sl ff}}}
\newcommand{\xffVar}{x_{\mbox{\sl ffVar}}}
\newcommand{\xrr}{x_{\mbox{\sl rr}}}
\newcommand{\xrMax}{x_{\mbox{\sl rMax}}}
\newcommand{\xdP}{x_{\mbox{\sl dP}}}
\newcommand{\xP}{x_{\mbox{\sl P}}}
\newcommand{\yfx}{y_{\mbox{\sl fx}}}

\usepackage{natbib}
\begin{document}

\title{Forecast verification for extreme value distributions with an application to probabilistic peak wind prediction}

\author{Petra Friederichs\footnote{\noindent \textit{Corresponding author address:} 
\newline{Petra Friederichs, Meteorological Institute, University of Bonn, Auf dem Huegel 20, 53121 Bonn, Germany}
\newline{E-mail: pfried@uni-bonn.de}} \\
\textit{\small{Meteorological Institute, University of Bonn, Germany}}
\and Thordis L. Thorarinsdottir \\
\textit{\small{Norwegian Computing Center, Oslo, Norway}}
}

\date{}

\maketitle

\begin{spacing}{1}

\begin{abstract}
\noindent
Predictions of the uncertainty associated with extreme events are a vital component of any prediction system for such events.  Consequently, the prediction system ought to be probabilistic in nature, with the predictions taking the form of probability distributions.  This paper concerns probabilistic prediction systems where the data is assumed to follow either a generalized extreme value distribution (GEV) or a generalized Pareto distribution (GPD).  In this setting, the properties of proper scoring rules which facilitate the assessment of the prediction uncertainty are investigated and closed-from expressions for the continuous ranked probability score (CRPS) are provided.  In an application to peak wind prediction, the predictive performance of a GEV model under maximum likelihood estimation, optimum score estimation with the CRPS, and a Bayesian framework are compared.  The Bayesian inference yields the highest overall prediction skill and is shown to be a valuable tool for covariate selection, while the predictions obtained under optimum CRPS estimation are the sharpest and give the best performance for high thresholds and quantiles.  

{\em Keywords:} Bayesian variable selection, continuous ranked probability score, extreme events, optimum score estimation, prediction uncertainty, wind gusts
\end{abstract}

\section{Introduction}

Extreme events in weather and climate such as high wind speeds, heavy precipitation or extremal temperatures are commonly associated with high impacts on both environment and society.  However, the physical processes leading to the extremes are usually generated on small scales and their prediction is contaminated by large uncertainty.  The need to determine uncertainties in the predictions of extreme events is stressed in the report from a recent workshop on extreme events in climate and weather \citep{Guttorp10}.  A prediction system for such events should therefore be probabilistic in nature, allowing for an assessment of the associated uncertainty \citep{Dawid84, Gneiting08}.  

The verification methods applied to these systems should thus necessarily be equipped to also handle the verification of uncertainty estimates.  \cite{Murphy93} argues that a general prediction system should strive to perform well on three types of goodness: there should be consistency between the forecaster's judgment and the forecast, there should be correspondence between the forecast and the observation, and the forecast should be informative for the user.  On a similar note, \cite{Gneiting07b} state that the goal of probabilistic forecasting should be to maximize the sharpness of the predictive distribution subject to calibration.  Here, calibration refers to the statistical consistency between the predictive distribution and the observation, while sharpness refers to the concentration of the predictive distribution; the sharper the forecast, the higher information value will it provide. The prediction goal of \cite{Gneiting07b} is thus equivalent to Murphy's second and third type of goodness. 

Verification methods that aim to attain these goals have been extensively studied in the literature, see e.g. \citet[Chapter 8]{Wilks11} for an excellent overview.  In this paper, we focus on the prediction of extreme events and our main objective is to assess the characteristics of proper scoring rules for extreme value distributions.  The framework of proper scoring rules can also be used for the parameter estimation in that a scoring rule is optimized over the training data, see e.g. \citet{Gneiting05b}.  Optimum score estimation returns unbiased parameter estimates \citep{Dawid07} and for the ignorance score, this equals maximum likelihood estimation for independent observations.  The continuous ranked probability score or CRPS \citep{Unger85,Hersbach00,Gneiting07} is of particular interest in our context, as it simultaneously assesses all of Murphy's types of goodness. A closed form expression of the CRPS has been calculated for a normal distribution \citep{Gneiting05b}, for a mixture of normals \citep{Grimit06}, for a truncated normal distribution \citep{Gneiting06}, and for the three parameter two-piece normal distribution \citep{Thorarinsdottir10a}. We derive closed-form expressions for the CRPS for the generalized extreme value distribution (GEV) and the generalized Pareto distribution (GDP).  

In an application to peak wind prediction, we compare minimum CRPS estimation, maximum likelihood estimation, and a Bayesian approach under the GEV using predictive performance as the comparison criteria.  Peak winds of short duration (few seconds) are a major cause of wind-related damage and crucial in wind hazard studies. Observations of peak winds are, however, sparse since only a small proportion of the weather stations provide peak wind speed observations.  Using data from the observation station Valkenburg in the Netherlands, we investigate how observations of other weather variables, including mean wind speed, precipitation, and pressure, may be used as covariates to obtain a predictive distribution for the peak wind speed. Furthermore, we show how a Bayesian regression variable selection method can simplify the covariate selection procedure when the space of potential models is large.  

The remainder of the paper is organized as follows.  In section~\ref{Sec:Methods}, we discuss how the predictive performance of probabilistic forecasts for extreme events may be assessed and provide the closed form expressions for the CRPS under the GEV and the GPD.  A detailed description of the derivation is given in the appendix.  Our case study on probabilistic forecasting and forecast verification for peak wind speed is presented in section~\ref{Sec:Application}. The paper closes with a discussion in section~\ref{Sec:Discussion}.

\section{Assessing predictive performance}\label{Sec:Methods}

\citet{Murphy87} and \citet{Murphy93} define a general framework for deterministic forecast verification based on two aspects of prediction quality: reliability and resolution.  Reliability, or calibration, relates to the correspondence between the observation and the forecast.  In our probabilistic setting, a forecast $F$ is perfectly calibrated if the conditional distribution of the observation $Y$ given the forecast is equal to $F$.  Resolution is a measure of information content which is closely related to the sharpness of the forecast.  In our setting, sharpness is defined with respect to the predictive distribution, where a sharp predictive distribution reflects high information content or entropy.  Note that sharpness has to be conditioned on $F$ being calibrated for it to be related to the resolution -- otherwise it is merely an attribute of the forecast distribution.  In the following, we discuss various methods to assess the calibration, or the reliability, and the resolution of a predictive distribution for data assumed to follow an extreme value distribution as in (\ref{Eq:GEV}) or (\ref{Eq:GPD}) below.

\subsection{Calibration}\label{Sec:Calibration}
 Let $F_1,\ldots,F_n$ denote the respective predictive distributions for the observations $y_1,\ldots,y_n$.  A standard method to assess the calibration of the forecasts is to apply the probability integral transform (PIT) \citep{Dawid84}.  The PIT is given by the value of the predictive distribution at the observation, $F_i(y_i)$.  If the forecasts are calibrated, the sample $\{F_i(y_i)\}_{i=1}^n$ should follow the standard uniform distribution.  This may e.g. be assessed graphically through a histogram, where a $\cup$-shaped histogram will indicate that the forecasts are underdispersive, while a $\cap$-shaped histogram will indicate that the forecasts are overdispersive, see e.g. \citet{Wilks11}.  The PIT histogram is the continuous counterpart of the verification rank histogram or the Talagrand diagram \citep{Anderson96,Hamill97}.   

The PIT values might also be displayed in terms of a PP-plot, where they are compared to $n$ equally spaced percentiles of the standard uniform distribution.  In extreme value theory, a very popular assessment of the calibration of a non-stationary extreme value model is the so-called residual quantile plot \citep{Coles01}.  Residual quantile plots emphasize potential deviations in the upper tail of the distribution. To this end, the PIT values and the uniform percentiles are transformed to quantiles through some inverse distribution function $G^{-1}$. In general, the inverse Gumbel is used for $G^{-1}$, as it is the standard reference distribution for models of block maxima. For threshold models an exponential distribution is often used.  

The calibration might also be estimated over a restricted range of forecasts, for instance situations with high forecast probability of extreme events.  A stratified PIT histogram or residual quantile plot would then be evaluated based on $\{F_i(y_i)\}_{i \in I}$ for some $I \subset \{1,\ldots,n\}$.  An important aspect here is that the subset $I$ has to be chosen independently of the observations $y_1,\ldots,y_n$, as the observed value is not known to the forecaster when a forecast is issued.   

\subsection{Scoring rules}

A variety of scores exist to assess the quality of a probabilistic forecast.  An important characteristic of a qualified scoring rule is its propriety \citep{Murphy73,Gneiting07,Broecker07}.  A scoring rule is (strictly) proper if the expected score for an observation $Y$ is optimized if (and only if) the true distribution of $Y$ is issued as the forecast.  Propriety will encourage honesty and prevent hedging which coincides with Murphy's first type of goodness \citep{Murphy93}.  We will only consider proper scoring rules in the following.  Furthermore, the scoring rules are negatively oriented such that a lower value means a better score.  

A widely used scoring rule is the logarithmic score proposed by \citet{Good52} which in meteorological applications is known under the name ignorance score \citep{Roulston02}. It is defined as 

\[
 S_{IGN}(f,y) = - \log(f(y) dy) .
\]
The ignorance score applies to the predictive density function $f$ and is proportional to the log-likelihood of the data with respect to the predictive density. Note, that $dy$ is ignored when calculating the log-likelihood. However, it must be taken into account if transformed variables are compared.  The associated expected score is the Shannon entropy, and the divergence function becomes the Kullback-Leibler divergence \citep{Gneiting07}.  It thus represents a score which is motivated by information theory.  Both \citet{Selten98} and \citet{Grimit06} have pointed out that, although the ignorance score is simple to calculate, it attributes a very strong penalty to events with low probability and is thus very sensitive to outliers.  

The continuous ranked probability score (CRPS) assesses the predictive skill of a forecast in terms of the entire predictive distribution $F$ \citep{Unger85,Hersbach00} and can thus assess both the calibration and the sharpness of the forecast simultaneously.  It is defined as \\

\begin{equation}\label{Eq:CRPS}
S_{CRP}(F,y) = \int_{-\infty}^{\infty} \left[ F(t) - H(t-y)\right]^2 dt
\end{equation}
and compares the forecast distribution $F$ and the empirical distribution of the observation $y$.  Here, $H(t-y)$ denotes the Heaviside step function using the half-maximum convention with $H(0)=0.5$. Note that an observation error might easily be introduced in the CRPS, e.g. assuming Gaussian errors and using a Gaussian distribution instead of the Heaviside step function.  

The CRPS can be decomposed into a reliability and a resolution part \citep{Hersbach00,Gneiting07}.  That is, we can write

\begin{equation}\label{Eq:CRPS Brier}
S_{CRP}(F,y) = \E |X-y| - \frac{1}{2} \E |X-X'|,
\end{equation}
where $X$ and $X'$ are independent random variables with distribution $F$.  In principle, any (strictly) proper scoring rules may be decomposed into an uncertainty, a reliability and, a resolution part \citep{Gneiting07, Broecker09}.  This representation of the CRPS is especially useful if a closed form expression of the CRPS is not available for $F$.  Let $\mathbf{x}$ and $\mathbf{x}'$ denote two independent samples of size $m$ from the predictive distribution $F$.  The representation in (\ref{Eq:CRPS Brier}) can then easily be approximated by

\begin{equation}\label{Eq:CRPS approx}
S_{CRP}(F,y) \approx \sum_{i=1}^m |x_i - y| - \frac{1}{2} \sum_{i=1}^m |x_i - x'_i|.
\end{equation}

Even though the forecast is given by a full predictive distribution $F$, it is often of interest to focus on the prediction of certain events, such as the probability of threshold exceedance.  We can assess the forecasters ability to predict events over a given threshold $u$ with the Brier score,

\begin{equation}\label{Eq:brier score}
S_B^u(F,y) = (p_u - \mathbbm{1}\{y \geq u\})^2,
\end{equation}
where $p_u = 1 - F(u)$ is the predicted probability of the realized value being greater or equal to the threshold $u$ \citep{Brier50}.  Note that the CRPS in (\ref{Eq:CRPS}) represents an integral of the Brier score over all possible thresholds.

Similarly, we might want to focus on the predictive performance in the upper tail.  This can be achieved by using the quantile score

\begin{equation}\label{Eq:quantile score}
 S^{\tau}_{Q}(F,y) = \rho_\tau(y-F^{-1}(\tau)),
\end{equation}
where $y$ is the event that materializes, $\tau$ is the quantile of interest, $\rho_\tau(u) = \tau u$ if $u\geq 0$, and $\rho_\tau(u) = (\tau-1) u$ otherwise \citep{Gneiting07,Friederichs07}.  A third alternative deviation of the CRPS is obtained using the quantile score, 

\begin{equation}\label{Eq:CRPS quantile}
 S_{CRP}(F,y)  = 2 \int_0^1 \rho_\tau(y-F^{-1}(\tau)) d\tau,
\end{equation}
see \citet{Laio07}, \citet{Gneiting07}, and \citet{Gneiting11}. 

When comparing various forecasters, it might often be advantageous to compare skill scores rather than the scores themselves \citep{Murphy73,Murphy74}.  A skill score measures the relative gain of a forecast with respect to the reference forecast and it is defined as 
\[
SS(F,y) = \frac{S(F,y)-S(F_{ref},y)}{S(F_{perfect},y)-S(F_{ref},y)} 
\]  
where $F^{ref}$ is a reference forecast used for all forecasters, and $F_{perfect}$ is the perfect forecast. In the case of CRPS and ignorance score the respective score of a perfect forecast is zero. A zero skill score represents no gain in predictive skill, while a perfect forecast would have a skill score of 100\%.  This approach is especially useful when comparing very high quantiles under the quantile score in (\ref{Eq:quantile score}) or very high thresholds under the Brier score in (\ref{Eq:brier score}) as these will be based on relatively small amount of data.

\subsection{Scoring rules for extreme value distributions}

Events are generally classified as being extreme by one of two criteria: the extremes are either given by a block maxima or they are defined as all values that exceed a given threshold.  The asymptotic behavior of the first type can be modeled by the generalized extreme value distribution (GEV),

\begin{equation}\label{Eq:GEV}
  F_{GEV}(y) =  \left\{ \begin{array}{lc}
    \exp\big(-(1+\xi\frac{y-\mu}{\sigma})^{-1/\xi}\big) ,&  \xi \ne 0 \\[0.2cm]
    \exp\big(-\exp(-\frac{y-\mu}{\sigma})\big) ,& \xi = 0  \end{array} \right. ,
\end{equation}
where $1 + \xi \frac{y-\mu}{\sigma} > 0$ for $\xi \ne 0$ \citep{Coles01,Beirlant04}.  The three parameters of the GEV are location $\mu$, scale $\sigma$ and shape $\xi$.  For threshold excesses, the generalized Pareto distribution (GPD) is usually applied.  It is given by  

\begin{equation}\label{Eq:GPD}
  F_{GPD}(y) = \left\{ \begin{array}{lc}
      1-(1+\xi \frac{y-u}{\sigma_u})^{-1/\xi},&  \xi \ne 0 \\
      1-\exp(-\frac{y-u}{\sigma_u}),& \xi = 0  \end{array} \right. ,
\end{equation}
where $u$ is a sufficiently large threshold, $\sigma_u$ is the scale and $\xi$ the shape parameter.

For $\xi \neq 0$, a closed form expression of the CRPS for the GEV is given by  

\begin{align}\label{Eq:CRPSGEVfinal}
  S_{CRP}(F_{GEV_{\xi \ne 0}},y) 
  = & \Big[  \mu - y - \frac{\sigma}{\xi} \Big]  \Big[ 1-2F_{GEV_{\xi \ne 0}}(y)\Big] \nonumber\\
  &- \frac{\sigma}{\xi} \Big[  2^{\xi} \Gamma(1-\xi)  - 2 \Gamma_l\big(1-\xi,-\log F_{GEV_{\xi \ne 0}}(y) \big) \Big],
\end{align}
where  $\Gamma$ denotes the gamma function and $\Gamma_l$ the lower incomplete gamma function.  For $\xi = 0$, this expression reads

\begin{equation}\label{Eq:CRPSGEV0final}
  S_{CRP}(F_{GEV_{\xi = 0}},y) =  \mu - y + \sigma [C-\log2] - 2 \sigma
  Ei\big(\log  F_{GEV_{\xi = 0}}(y) \big),
\end{equation}
where $C \approx 0.5772$ is the Euler-Mascheroni constant and $Ei(x) = \int_{-\infty}^x \frac{e^t}{t}  dt$ is the exponential integral \citep{Abramowitz64}.  The derivations are given in Appendix A.  In the case of the GPD, we get 

\begin{align}\label{Eq:CRPS_GPD}
  S_{CRP}(F_{GPD_{\xi \ne 0}},y)  
  = & \Big[ u - y - \frac{\sigma_u}{\xi}\Big] \Big[ 1 - 2 F_{GPD_{\xi \ne 0}}(y) \Big] \nonumber\\
  &- \frac{2\sigma_u}{\xi (\xi-1)} \Big[ \frac{1}{\xi-2} + \Big[1-F_{GPD_{\xi \ne 0}}(y)\Big]\Big[1+\xi\frac{y-u}{\sigma_u} \Big]\Big]
\end{align}
for $\xi \ne 0$ and 

\begin{equation}\label{Eq:CRPS_GPD0}
  S_{CRP}(F_{GPD_{\xi = 0}},y) = y-u -  \sigma_u\Big[2 F_{GPD_{\xi = 0}}(y) -\frac{1}{2}\Big]
\end{equation}
for $\xi=0$.  The derivations for a GPD are given in Appendix B.

\begin{figure}[t]
\begin{center}
\includegraphics[width=\textwidth]{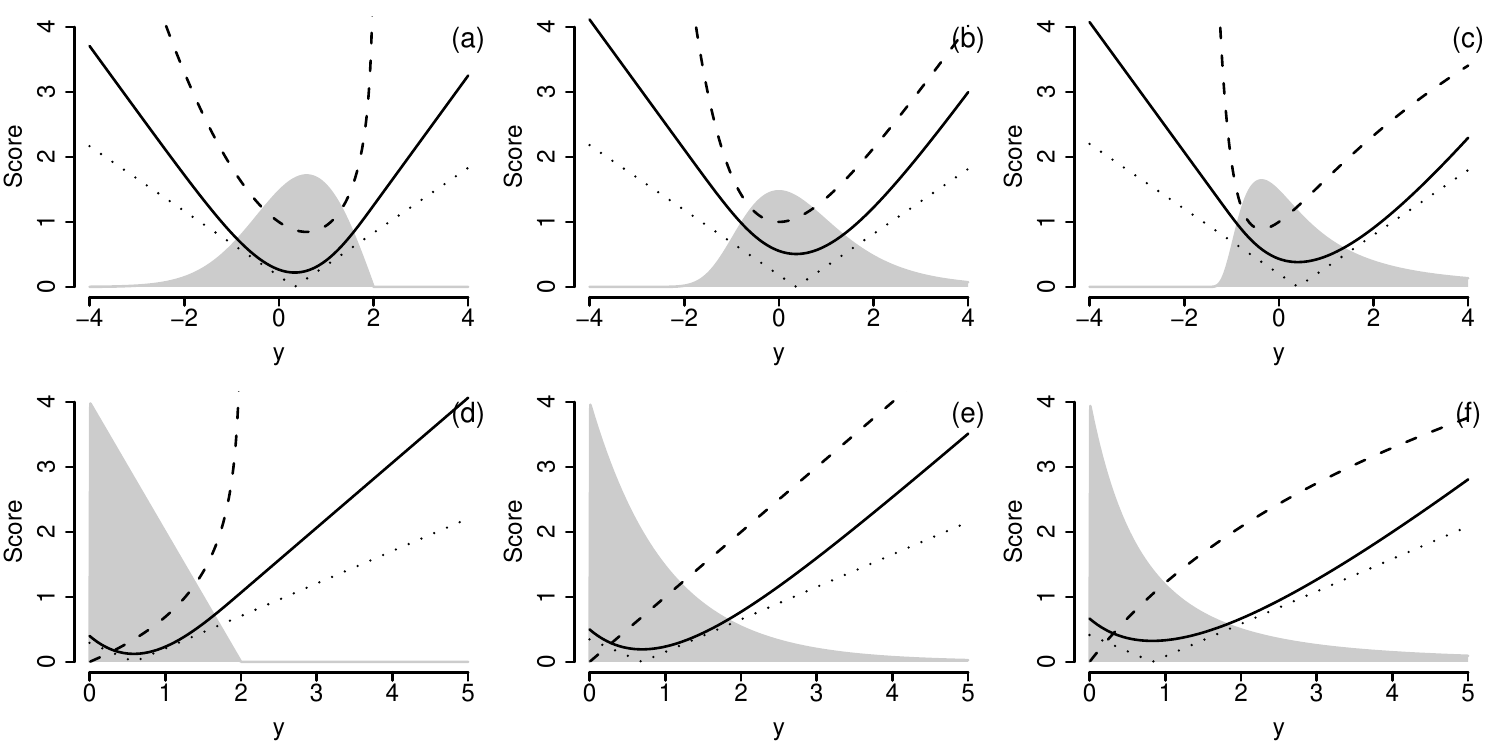}
\caption{The CRPS (solid lines), the ignorance score (dashed lines), and the quantile score for $\tau = 0.5$ (dotted lines) as a function of the observation value for the GEV (top row) and the GPD (bottom row).  The predictive distributions have zero location ($\mu = 0$), standard scale ($\sigma = 1$), and shape $\xi = -0.5$ (first column), $\xi = 0$ (second column), or $\xi = 0.5$ (third column).  The corresponding predictive densities are indicated in gray.}
\label{Fig:CRPS}
\end{center}
\end{figure}

Figure \ref{Fig:CRPS} shows the CRPS, the ignorance score and the quantile score for $\tau = 0.5$ for the GEV with shape parameter $\xi=-0.5$ (Weibull type), $\xi=0$ (Gumbel type), and $\xi=0.5$ (Fr\'echet type), as well as for the GDP with the same parameter values.  The ignorance score has a sharper minimum than the other scores and takes its minimum at the mode of the predictive distribution while the other two scores take their minimum at the median of the distribution.  For the CRPS under the GEV, this can easily be shown by introducing the median, $F_{GEV}(0.5) = \mu+\frac{\sigma}{\xi}( (\log 2)^{-\xi}-1)$ for $\xi \neq 0$ and $F_{GEV}(0.5) = \mu -\sigma \log( \log 2)$ for $\xi =0$, into the derivative of the score with respect to $y$.

\section{Predicting peak wind speed}\label{Sec:Application}

We apply the verification measures discussed in the previous sections to predictions of daily peak wind speed at a location in the Netherlands over a period from January 1, 2001 to July 1, 2009.  The goal is to derive a prediction model for daily peak wind speed using the observations of other meteorological variables as covariates with out-of-sample data used for the parameter estimation. In our application, the covariates are observed at the same time and location as the observation.  We thus derive nowcasts rather than real forecasts in time for daily peak wind speed.  In view of the sparse observational network for gust observations, such an analysis can be very useful, e.g. for forecast verification of wind gust warnings \citep{Friederichs09}.  Furthermore, the covariates might easily be replaced by the corresponding outputs from numerical weather prediction (NWP) models, thereby providing a probabilistic methodology for model output statistics, as NWP models only give diagnostic estimates for peak wind speed.

For the modeling, we use generalized extreme value distributions under both a Bayesian forecasting framework and a frequentist framework where the parameters are estimated by minimizing a proper scoring rule.  That is, we assume that peak wind speed observations follow a non-stationary GEV (\ref{Eq:GEV}) where the parameters depend on covariates $\mathbf{x}$ through functions of the form
\begin{eqnarray}\label{Eq:nsGEV}
 \mu(\mathbf{x}) = \mu_0 + \mu_1 x_1 + \mu_{2} x_{2} + \ldots, \quad
 h( \sigma(\mathbf{x})) = \sigma_0 + \sigma_1 x_1 +  \sigma_{2} x_{2} + \ldots, \quad
 \xi = \xi_0,
\end{eqnarray}
where $h(\cdot)$ is a link function in analogy to generalized linear models \citep{Fahrmeir01}.
{The shape parameter is very sensitive to sampling uncertainty. Including covariates to the shape parameter largely increases the uncertainty not only of the shape parameters estimate itself, but also of the other parameters \citep{Friederichs09}. The benefit in turn is generally small. For this reason, the shape parameter is kept independent of covariates. However, in a forecasting procedure, alike the other parameters, the shape parameter is estimated on a training data set and used to provide out-of-sample prediction. This out-of-sample prediction and verification is realized by a cross-validation procedure (see section \ref{Sec:MS}).}

As an alternative forecast, we apply a standard approach in meteorology and wind engineering, a generalized linear model \citep[GLM, e.g.,][]{McCullagh99} for the gust factor.  The gust factor $G$ is defined as 

\begin{equation}
 G = \frac{y_{\fx}}{x_{\ff}} - 1,
\end{equation}
where $x_{\ff}$ denotes the mean wind speed observation and $y_{\fx}$ the peak wind speed observation.  The gust factor is generally assumed to follow a lognormal distribution, while the most simple approaches assume a constant gust factor.  We base our approach on the work of \citet{Weggel99} and \citet{Jungo02} and use a GLM for lognormal residuals.  This corresponds to a standard linear regression model for $\log G$ with the expected value being modeled as 

\begin{equation}
 \E[ \log G ] = \beta_0 +  \beta_1 x_1 + \beta_2  x_{2} + \ldots.
\end{equation}
The GLM is estimated using iteratively reweighted least squares as provided by the \verb glm  function in {\tt R} \citep{R}.

\subsection{Data}\label{Sec:Data}
The Royal Netherlands Meteorological Institute freely provides daily weather data from numerous observation locations in the Netherlands\footnote{\url{http://www.knmi.nl/climatology/daily_data/download.html}}.  The meteorological station data contain observations of temperature, sunshine, cloud cover and visibility, air pressure, precipitation, mean wind, and peak wind speed for the specified station.  We have chosen the station number 210, Valkenburg, with hourly observations from January 1, 2001 to July 1, 2009 which gives us a total of $3104$ daily observations.  The daily observed peak wind $\yfx$ is measured as a 24-hour (00UTC-00UTC) maximum over the observed hourly wind gusts.  The gust observations are measured in 1 m/s which makes the data quasi discrete.  In order to obtain a more continuous spectrum of values, we added a uniformly distributed noise to the gust observations.  The hourly mean wind speed is given by the mean wind speed during the 10-minute period preceding the time of observation.  We process these data to daily mean wind $\xff$ and daily wind variance $\xffVar$ by taking the average and the variance, respectively, over the 24 hourly observations as above.  A similar procedure yields the mean rain rate $\xrr$ and the maximum rain rate $\xrMax$ over 24 hours.  Finally, we also consider the mean 24-hour pressure $\xP$ and the 24-hour pressure tendency $\xdP$ as potential covariates. The pressure tendency $\xdP$ is estimated by fitting a linear trend function (i.e., linear function in time) to the 24 hourly pressure observations. The slope estimate yields an estimate of $\xdP$. Note that the covariates are normalized before entering as a covariate.  Normalization is not necessary but makes the inference procedures more stable.   

\begin{figure}[t]
\begin{center}
\includegraphics[width=0.8\textwidth,angle=0]{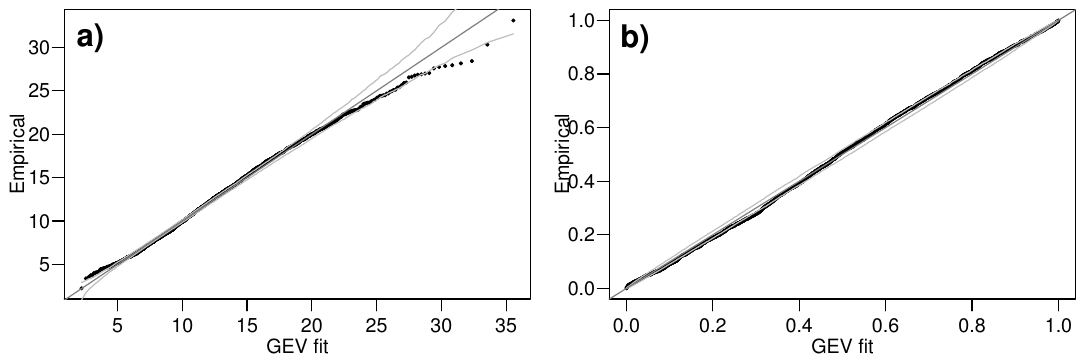}
\caption
[]
{The goodness of fit of a seasonal GEV model applied to 24h peak wind speed observations at Valkenburg, the Netherlands from January 1, 2001 to July 1, 2009 measured by a QQ-plot  (a) and a PP-plot (b).  The dark gray line represents the seasonal GEV model estimates, the light gray lines indicate the 95\% confidence interval derived by parametric bootstrapping, and the black dots denote the empirical estimates. 
}
\label{Fig:GOF}
\end{center}
\end{figure}

Peak wind speed refers to the maximum wind speed in a given time period.  In operational terms, this refers to the highest 3-second average value recorded in a particular time period (e.g. 1 hour).  Consequently, a block maxima approach using a GEV represents a natural approach to model the peak wind speed observations.  
Figure \ref{Fig:GOF} represents a goodness of fit for a GEV model fitted to the daily peak wind speed observations in our data set using maximum likelihood estimation.  
In order to account for seasonal variations, we included the annual cycle in terms of a annual sine and cosine function as covariate to the location and scale parameter.
Both parameters exhibit significant annual changes. The relation of the shape parameter to the annual cycle is not significantly different from zero, neither any relation to a half-annual sine or cosine function. Although significant, the annual cycle accounts for only about 5\% of the variance in the daily peak wind speed observations.  The seasonally varying GEV exhibits a shape parameter of about $\hat \xi = -0.022$ with a standard error of $ 0.014$ and provides an overall acceptable fit even though it seems to slightly overestimate high quantiles, see Figure~\ref{Fig:GOF}(a).

\subsection{Optimum score estimation}

In our frequentist approach, we apply both classic maximum likelihood estimation (GEV-MLE) and minimum CRPS estimation (GEV-CRPS) for the parameter estimation of the predictive distribution function.  For the minimum CRPS estimation, we minimize the function 

\[
\sum_i S_{CRP}(F_i,y_{\fx,i})
\]
where the index runs over the training set, $S_{CRP}$ is given by either (\ref{Eq:CRPSGEVfinal}) or (\ref{Eq:CRPSGEV0final}) depending on the value of the shape parameter $\xi$, and $F_i$ is the predictive distribution for a covariate vector $\mathbf{x}_i$.  The predictive distribution for $Y$ is a GEV with

\begin{equation}\label{Eq:GEV-MLE}
 P(Y\leq y| \mathbf{x}) = F_{GEV}(y;\mu(\mathbf{x}),\sigma(\mathbf{x}),\xi),
\end{equation}
where the parameters $\mu(\mathbf{x}), \sigma(\mathbf{x})$, and $\xi$ are given by (\ref{Eq:nsGEV}).  Here, we apply both the identity and the logarithmic link function for the scale parameter $\sigma$.  The latter ensures a positive scale parameter and guarantees valid predictions.  

Numerical optimization was first performed using the quasi-Newton Broyden-Fletcher-Goldfarb-Shanno (BFGS) and the Nelder-Mead methods implemented in {\tt R} \citep{R} in parallel and the best estimates in terms of optimum score were used. However, as we included more covariates, the optimization became less stable. For that reason, all optimum score estimates are derived using a Simulated Annealing algorithm (SANN). The SANN method in {\tt R} by default uses a variant of Simulated Annealing given in \citet{Belisle92} and is useful for getting good estimates on rough surfaces. Although the SANN algorithm is slower than the BFGS and the Nelder-Mead, it provides more robust estimates.

\begin{figure}[t]
\begin{center}
\includegraphics[width=0.8\textwidth,angle=0]{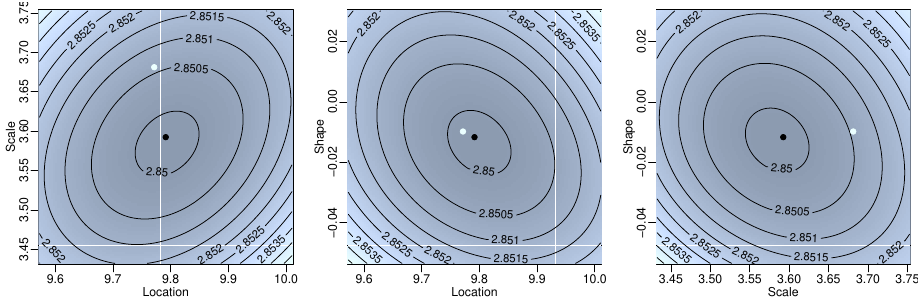}
\caption
{The in-sample ignorance score for the stationary GEV over the full data set.  
The plots show changes in the score when the location $\mu_0$, scale $\sigma_0$, and shape $\xi_0$ are varied pairwise while the remaining parameter is fixed at the MLE.  The black dots indicate the MLE and the white dots the minimum CRPS estimate. 
}
\label{Fig:IGNGEV}
\end{center}
\end{figure}

\begin{figure}[tb]
\begin{center}
\includegraphics[width=0.8\textwidth,angle=0]{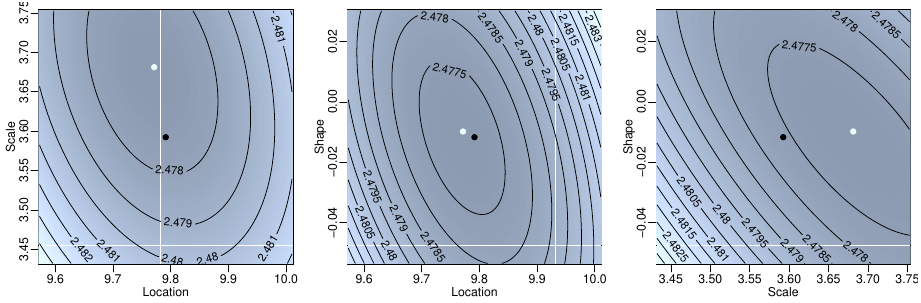}
\caption
{The in-sample CRPS for the stationary GEV over the full data set.  
The plots show changes in the score when the location $\mu_0$, scale $\sigma_0$, and shape $\xi_0$ are varied pairwise while the remaining parameter is fixed at the minimum CRPS estimate.  
The white dots indicate the minimum CRPS estimates and the black dots the MLE. 
}
\label{Fig:CRPSGEV}
\end{center}
\end{figure}

In order to compare the maximum likelihood and the minimum CRPS estimation, we investigate two-dimensional surfaces of the in-sample ignorance score and the CRPS over the entire data set.
The link function for the scale parameter $\sigma$ is the identity.  Figure~\ref{Fig:IGNGEV} shows two-dimensional surfaces of the log-likelihood function, or the ignorance score, for pairwise combinations of $\mu_0$, $\sigma_0$, and $\xi_0$.  In each plot, the other parameter is fixed at its respective maximum likelihood estimate (MLE).  The log-likelihood function is displayed over a range that covers the parameter estimate plus-minus three standard errors as derived from the profile likelihood \citep{Coles01}.  Similar two-dimensional surfaces are displayed for the CRPS in Figure~\ref{Fig:CRPSGEV}, where in each plot, the other parameter is fixed at its respective minimum CRPS estimates.  The CRPS is displayed over the same parameter range as the log-likelihood function in Figure~\ref{Fig:IGNGEV}.  The MLE and the minimum CRPS estimates differ only slightly, or by less than two standard errors for all parameters (cf. the black and white dots in Figure~\ref{Fig:IGNGEV} and \ref{Fig:CRPSGEV}). 
The pairwise dependence of changes in the parameter values on the function surfaces is quite different for the two functions. It is particularly strong for the CRPS where the dependence is negatively oriented. Further, the CRPS function is less sharp than the log-likelihood function in the direction of $\xi_0$.

\subsection{Bayesian forecasting framework}\label{Sec:Bayesian}

Let $l(\mathbf{y}_{\fx}|\bm\theta,\mathbf{X})$ denote the likelihood function under the GEV model in (\ref{Eq:GEV}), where $\mathbf{y}_{\fx}$ are the observed peak winds over the training set, $\mathbf{X}$ is the matrix of corresponding covariates, and $\bm\theta$ is a vector of the model parameters in (\ref{Eq:nsGEV}) with a logarithmic link function for the scale parameter.  Given a prior distribution $f(\bm\theta)$ for $\bm\theta$, the joint posterior distribution of the parameters given the data is 

\[
f(\bm\theta|\mathbf{y}_{\fx},\mathbf{X}) \propto l(\mathbf{y}_{\fx}|\bm\theta,\mathbf{X}) f(\bm\theta). 
\] 
The predictive density for a new observation $y'$ with a covariate vector $\mathbf{x}'$ is given by

\begin{equation}\label{Eq:Bayes pred}
f(y'|\mathbf{x}',\mathbf{y}_{\fx},\mathbf{X}) = \int l(y'|\bm\theta,\mathbf{x}') f(\bm\theta|\mathbf{y}_{\fx},\mathbf{X}) \dd \bm\theta.
\end{equation}
While this density is generally not a GEV density, it has the advantage that it reflects uncertainty both in the model and in the future observations, see e.g. \cite{Coles01}.  For the GEV model, the integral in (\ref{Eq:Bayes pred}) is usually intractable.  However, the density may easily be approximated by a large sample from the predictive distribution, see \citet[section 4.3]{Hoff09}.

We use non-informative independent normal priors for the parameters and set $\mu_j, \sigma_j \sim N(0,10^4)$ for all $j$, and $\xi_0 \sim N(0,10^2)$.  For stationary GEV distributions, it is common to use the logarithmic link function for the scale parameter, see e.g. \citet{Stephenson04} and \citet{Galiatsatou08}.  We follow their practice and only use the logarithmic link function for this part of the analysis.  Our priors correspond to setting $\mu_0,\log(\sigma_0) \sim N(0,10^4)$ and $\xi_0 \sim N(0,10^2)$ in the stationary case.  \citet{Stephenson04} and \citet{Galiatsatou08} argue that, in the stationary case, a trivariate normal distribution would be preferred here, as a negative dependence between $\sigma_0$ and $\xi_0$ is  a priori expected.  As our inference is based on a relatively large data set, we refrain from this given the computational complexity that such a prior would induce in the non-stationary case.  We then apply a Metropolis within Gibbs algorithm to obtain samples from the marginal posterior distributions using normal distributions with a small variance as proposal distributions for the updates.  For each parameter estimation, we run 100.000 iterations of the Metropolis within Gibbs algorithm with a burn-in period of 25.000 iterations.  Classical diagnostics such as traceplots and running mean plots (not shown) show good mixing and indicate that this is sufficient for convergence.

In this case study, we consider six possible covariates and each of these can influence both the location and the scale parameter.  We thus have a total of $2^{12}$ possible models in our model space.  Besides considering specific models, we also perform a variable selection procedure over the entire model space.  That is, we write each regression coefficient $\mu_j$ for $j \geq 1$  as $\mu_j = z_j \nu_j$, where $z_j \in \{0,1\}$ and $\nu_j \in \real$ such that the $z_j$'s indicate which regression coefficients are non-zero.  The regression equation for $\mu$ becomes

\[
\mu(\mathbf{x}) = \mu_0 + z_1 \nu_1 \xff + z_2 \nu_2 \xffVar + z_3 \nu_3 \xrr + z_4 \nu_4 \xrMax + z_5 \nu_5 \xP + z_6 \nu_6 \xdP, 
\]
and similarly for $\log(\sigma(\mathbf{x}))$.  Each model indication parameter $z_j$ is a priori equal to either $0$ or $1$ with probability $1/2$, while the regression parameters $\nu_j$ have independent $N(0,10^4)$ priors as before.  Again, the parameters are updated iteratively using a Metropolis within Gibbs updating scheme.  

Our algorithm is equivalent to a reversible jump MCMC algorithm \citep{Green95} and the posterior inclusion probability of a covariate $x$ equals the average value of the corresponding indicator variable $z$ over the posterior sample.  A related Bayesian inference framework is proposed in \citet{ElAdlouni09} where the authors suggest a birth-death MCMC procedure for covariate selection in generalized extreme value models and apply their framework to annual maximum precipitation data.  However, the birth-death MCMC algorithm is quite complex to implement compared to the fairly simple regression variable selection method described in \citet{Hoff09} which we have applied.  For more details on the regression variable selection algorithm and Bayesian inference in general, see \cite{Hoff09}.     

\subsection{Model selection}\label{Sec:MS}

In order to assess the predictive performance of our prediction approaches, we pursue a common approach in atmospheric science and use an out-of-sample verification.  The prediction is performed in a cross-validation manner, in that we iteratively leave out one year of data, estimate the parameters of the statistical models based on the remaining data, and then perform an out-of-sample prediction.  In this way, independent predictions for the complete time series are generated and used for verification and model selection.  

\begin{table}[t]
  \centering
  \caption{Average posterior inclusion probabilities for the covariates in a Bayesian model selection framework over all years.  The probabilities are given in percentages.}
  \label{tab:inclusion prob}
  \smallskip
  \begin{tabular}{ld{3.1}d{3.1}}
    \toprule
    Variable & \multicolumn{1}{c}{Location} & \multicolumn{1}{c}{Scale} \\
    \midrule 
    Mean wind speed ($\ff$)     & 100.0 & 100.0 \\
    Wind variance ($\ffVar$)    & 100.0 & 100.0 \\
    Mean rain rate ($\rr$)      & 0.1 & 0.0 \\
    Maximum rain rate ($\rMax$) & 100.0 & 100.0 \\
    Pressure ($\slP$)             & 100.0 & 10.8 \\
    Pressure tendency ($\dP$)   & 100.0 & 0.0 \\
    \bottomrule  
  \end{tabular}
\end{table}

Average posterior inclusion probabilities for the covariates under the Bayesian regression variable selection framework are given in Table~\ref{tab:inclusion prob}.  Three covariates, mean wind speed ($\ff$), wind variance ($\ffVar$), and maximum rain rate ($\rMax$) have very high inclusion probabilities for both the location parameter $\mu$ and the scale parameter $\sigma$ in (\ref{Eq:nsGEV}).  In addition, pressure ($\slP$) and pressure tendency ($\dP$) also have high posterior inclusion probabilities for the location parameter $\mu$.  Note that the inclusion probabilities are calculated after the burn-in period has been removed.  Especially for the low inclusion probabilities, the chains show a high degree of mixing in the burn-in period.  Although seasonality may be captured by the existing covariates, we have further investigated adding annual cycles to the model.  The results indicate that a small improvement in predictive performance (i.e.\ of the order of 0.02)  may be obtained by including an annual cycle in the location parameter. Consequently, only a small fraction of seasonality is left unexplained when no annual cycle is concidered explicitly.  For clarity of exposition, we omit this factor in the following.

\begin{table}[t]
  \centering
  \caption{Average predictive performance of non-stationary peak wind speed predictions at Valkenburg from January 1, 2001 to July 1, 2009.  The performance is measured in terms of the continuous ranked probability score (CRPS), which is given in meters per second, and the ignorance score (IGN).  The covariates considered here are mean wind speed {\sl ff}, wind variance {\sl ffVar}, mean rain rate {\sl rr}, maximum rain rate {\sl rMax}, pressure {\sl P}, and pressure tendency {\sl dP}.  VS stands for Bayesian variable selection approach.  The link function $h(\sigma)$ used for the scale parameter  is indicated in parenthesis and the optimal score in each column is indicated in bold. }
  \label{tab:predictors}
  \smallskip
  \begin{tabular}{ld{1.2}d{1.2}d{1.2}d{1.2}d{1.2}d{1.2}d{1.2}}
    \toprule
    Covariate(s)           & \multicolumn{1}{c}{None} & \multicolumn{1}{c}{{\sl ff}} & \multicolumn{1}{c}{{\sl ff}} & \multicolumn{1}{c}{{\sl ff}}  & \multicolumn{1}{c}{{\sl ff}}  & \multicolumn{1}{c}{VS}\\
    & & & \multicolumn{1}{c}{{\sl ffVar}} & \multicolumn{1}{c}{{\sl ffVar}} & \multicolumn{1}{c}{{\sl ffVar}}  & \\
    & & & & \multicolumn{1}{c}{{\sl rMax}} & \multicolumn{1}{c}{{\sl rMax}} & \\ 
    & & & & & \multicolumn{1}{r}{{\sl P}$^{\textup{ $\mu$ only}}$} & \\
    & & & & & \multicolumn{1}{c}{\dP$^{\textup{ $\mu$ only}}$} & \\
    \midrule 
    CRPS (m/s)\\
    \midrule
    GEV-MLE (id)         & \mathbf2.\mathbf4\mathbf8 & \mathbf1.\mathbf0\mathbf7 & 0.86 & 0.82 & 0.80 & \multicolumn{1}{c}{---}\\
    GEV-MLE (log)        &      & \mathbf1.\mathbf0\mathbf7 & 0.86 & 0.82 & 0.80 & \multicolumn{1}{c}{---}\\
    GEV-CRPS (id)        & \mathbf2.\mathbf4\mathbf8 & \mathbf1.\mathbf0\mathbf7 & 0.86 & \mathbf0.\mathbf8\mathbf1 & \mathbf0.\mathbf7\mathbf9 & \multicolumn{1}{c}{---}\\
    GEV-CRPS (log)       &      & \mathbf1.\mathbf0\mathbf7 & 0.86 & 0.82 & 0.80 & \multicolumn{1}{c}{---}\\
    GEV-Bayes (log)      & 2.49 & 1.08 & \mathbf0.\mathbf8\mathbf4 & \mathbf0.\mathbf8\mathbf1 & \mathbf0.\mathbf7\mathbf9 & \mathbf0.\mathbf8\mathbf0 \\
    lognormal GLM        & 1.37 & 1.09 & 0.95 & 0.92 & 0.90 & \multicolumn{1}{c}{---}\\
    \midrule 
    IGN\\
    \midrule
    GEV-MLE (id)         & \mathbf2.\mathbf8\mathbf5 & \mathbf2.\mathbf0\mathbf0 & \mathbf1.\mathbf7\mathbf8 & \mathbf1.\mathbf7\mathbf3 & \mathbf1.\mathbf7\mathbf1 & \multicolumn{1}{c}{---}\\
    GEV-MLE (log)        &      & \mathbf2.\mathbf0\mathbf0 & \mathbf1.\mathbf7\mathbf8 & \mathbf1.\mathbf7\mathbf3 & 1.72 & \multicolumn{1}{c}{---}\\
    GEV-CRPS (id)        & \mathbf2.\mathbf8\mathbf5 & \mathbf2.\mathbf0\mathbf0 & \mathbf1.\mathbf7\mathbf8 & \mathbf1.\mathbf7\mathbf3 & 1.72 & \multicolumn{1}{c}{---}\\
    GEV-CRPS (log)       &      & \mathbf2.\mathbf0\mathbf0 & \mathbf1.\mathbf7\mathbf8 & 1.74 & 1.73 & \multicolumn{1}{c}{---}\\
    GEV-Bayes (log)      & \mathbf2.\mathbf8\mathbf5 & \mathbf2.\mathbf0\mathbf0  & \mathbf1.\mathbf7\mathbf8 & \mathbf1.\mathbf7\mathbf3 & \mathbf1.\mathbf7\mathbf1 & \mathbf1.\mathbf7\mathbf1 \\   
    lognormal GLM        & 2.20& 2.03 & 1.90 & 1.86 & 1.85 & \multicolumn{1}{c}{---}\\
    \bottomrule  
  \end{tabular}
\end{table}

Table~\ref{tab:predictors} shows the average CRPS and ignorance score over all days in the test set for all the prediction methods considered in this study and different sets of covariates for each method. To calculate the CRPS in Table~\ref{tab:predictors}, we have applied (\ref{Eq:CRPSGEVfinal}) and (\ref{Eq:CRPSGEV0final}) for the GEV models while for the Bayesian method, we have applied the approximation in (\ref{Eq:CRPS approx}).  Since the logarithm of the gust factor $log(G)$ follows a normal distribution, the CRPS of the lognormal GLM is calculated using the closed-form expression in \citet{Gneiting05b}.  To calculate the ignorance score for the Bayesian method, we obtain a non-parametric density estimate for the predictive density based on a large sample from the posterior predictive distribution.  The ignorance score for the lognormal GLM reads $S_{IGN}(f_{\cal N},\log(G)) = - \log(f_{\cal N}(\log(G))/(y-\ff) dy)$, where $y$ is the respective  gust value and $f_{\cal N}$ is the predictive normal distribution of $\log(G)$.  
{Note that the scores are calculated on cross-validated predictions (Sec. \ref{Sec:MS}), in order to provide an out-of-sample prediction and verification.
Further, we estimate the sampling uncertainty of the CRPS and ignorance score via the bootstrap method \citep{Efron93}.} To that end, we recalculated the scores by resampling the prediction-observation pairs using the bootstrap method with replacement. The standard deviations of the CRPS and ignorance score within a 1000 member bootstrap sample vary between $0.01$ and $0.03$ for all the GEV methods.

Including covariates in the prediction improves the predictive performance significantly compared to using a stationary model. The decrease in CRPS amounts to about 1.4 (1.7) when including {\sl ff} ({\sl ff, ffVar, rMax, P, dP}).  With respect to a sampling uncertainty of about $0.03$ this decrease is highly significant.  For the models compared here, all method show the highest predictive skill when only covariates with high posterior inclusion probability are included.  Including {\sl P} and {\sl dP} in the scale parameter and {\sl rr} in both the location and the scale parameter, or a subset of these, does not provide significant improvements.  Although the lognormal distribution seems to provide a better fit to the data in the stationary case, the non-stationary GEV models clearly outperform the lognormal GLM except if only {\sl ff} is used as covariate. The improvement of the CRPS amounts to about $0.1$ which is well above the sampling uncertainty of the CRPS. The differences between the various GEV approaches are more subtle and not very significant.  For the frequentist methods, the identity link for the scale parameter yields slightly higher skill than the logarithmic link, especially in terms of the ignorance score for the best non-stationary model.  For the more complex models, with three covariates or more, the optimum CRPS methods performs minimally better than the maximum likelihood methods in terms of the CRPS and the other way around for the ignorance score.  The best scores are generally obtained within the Bayesian framework.

\subsection{Predictive performance}\label{Sec:Performance}

Based on the results in the previous section, we now focus on the non-stationary GEV with location parameter

\begin{equation}\label{eq:optimal location}
\mu(\mathbf{x}) = \mu_0 + \mu_1 \xff + \mu_{2} \xffVar + \mu_3 \xrMax + \mu_4 \xP + \mu_5 \xdP
\end{equation}
and scale parameter

\begin{equation}\label{eq:optimal scale}
 h( \sigma(\mathbf{x})) = \sigma_0 + \sigma_1 \xff +  \sigma_{2} \xffVar + \sigma_3 \xrMax,
\end{equation}
where $h$ is the identity link for the optimum score estimation and the logarithm for the Bayesian method. {Although the differences are barely significant in table \ref{tab:predictors}, there is evidence that the identity link model provides slightly better skill compared to the logarithm when using optimum score estimation. However, the following results also apply to the logarithm link model.} For comparison, we also investigate the performance of the lognormal GLM method.  Verification is then based on a variety of diagnoses.  

\begin{figure}[tb]
\begin{center}
\includegraphics[width=0.75\textwidth,angle=0]{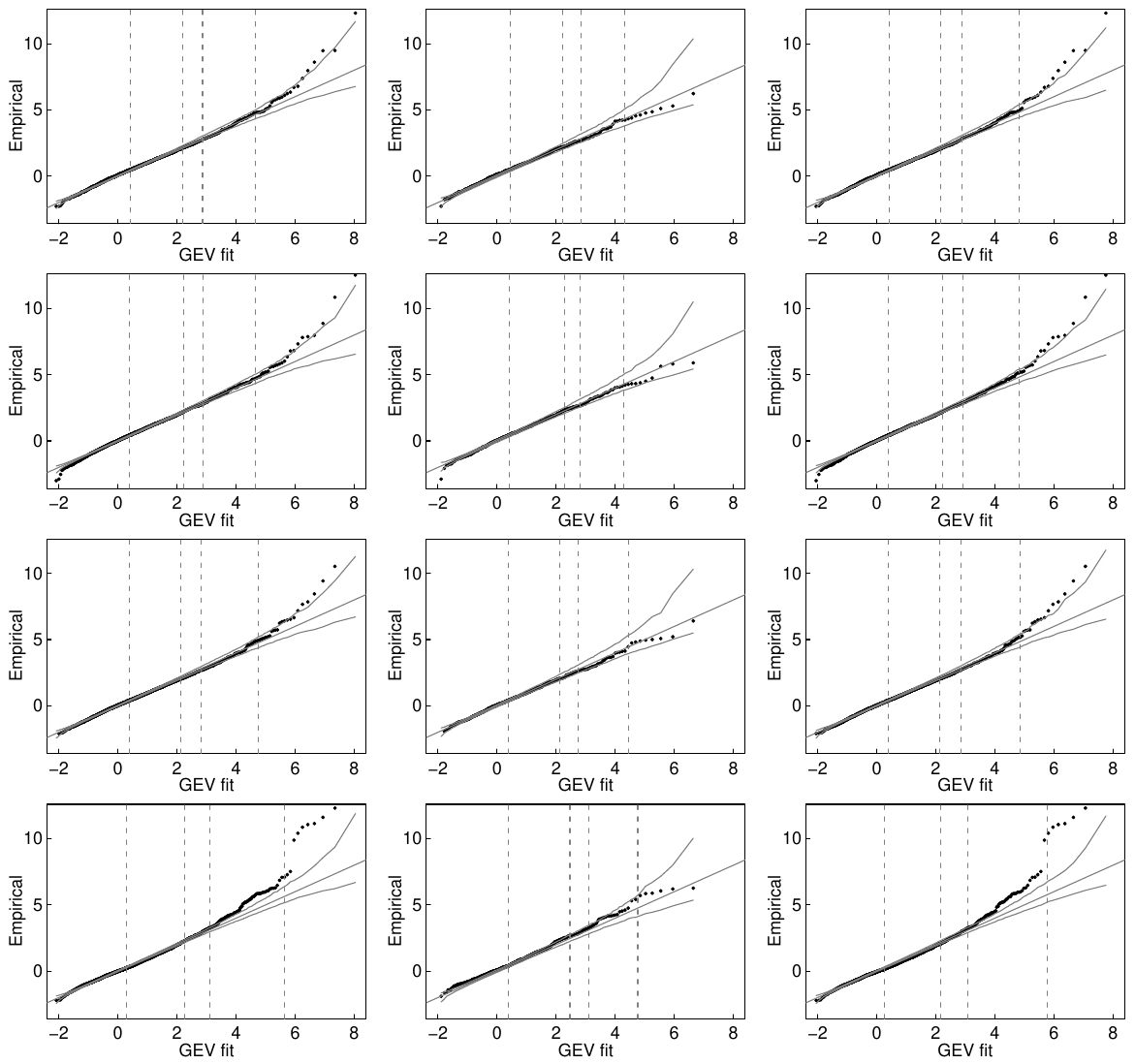}
\caption{Residual quantile plots on Gumbel scale for models with the location parameter in (\ref{eq:optimal location}) and the scale parameter in (\ref{eq:optimal scale}).  The prediction methods compared here are the GEV-MLE method with identity link (1st row), the GEV-CRPS method with identity link (2nd row), the Bayes method (3rd row), and the lognormal GLM (4th row).  The first column shows the QQ-plots over all forecasts, the second column shows forecasts with $\xff > q_{.75}$, and the third column shows forecasts with $\xff \leq q_{.75}$.  The vertical lines indicate (from left to right) the residual 0.5, 0.9, 0.95, and 0.99 quantiles.  
}
\label{Fig:resQQ}
\end{center}
\end{figure}

The calibration, or the reliability, of the predictions is assessed by using residual quantile plots based on the standard Gumbel distribution, see Figure~\ref{Fig:resQQ}.  The first column of Figure~\ref{Fig:resQQ} shows the residual quantile plots (section \ref{Sec:Calibration}) for the complete set of observations.  For the lowest 99\% of the residuals, the bulk of the residuals lies within the 95\% uncertainty band of the residual distribution for all three GEV methods.  However, the models show a significant underestimation of the highest 1\% of the residuals.  The underestimation is even more prominent for the logarithmic GLM where significant underestimation is observed for the highest 5\% of the residuals.  An additional stratification is based on the covariate $\xff$, where predictions with below normal mean wind speed ($\xff \leq q_{.75}$) and above normal mean wind speed ($\xff > q_{.75}$) are considered separately, where $q_{.75}$ is the $0.75$ quantile of the $\xff$ observations.  In situations where the mean wind $\xff$ is above its 0.75 quantile, all models provide calibrated predictive distributions (Fig.~\ref{Fig:resQQ}, second column).  The uncalibrated large residuals only occur during weak wind situations (Fig.~\ref{Fig:resQQ}, third column).  Since strong gusts are very unlikely during these weather situations, this miscalibration is probably not highly relevant for the gust prediction.

The three GEV methods in Figure~\ref{Fig:resQQ} thus all appear to be fairly well calibrated and we can compare the sharpness of the predictive distributions in concurrence with the forecasting principle of \citet{Gneiting07b}.  The GEV-CRPS method yields the sharpest predictions with an average predictive standard deviation of $1.43 \, (\pm 0.44)$, while this value is $1.48 \, (\pm 0.57)$ for the Bayesian method and $1.51 \, (\pm 0.48)$ for the GEV-MLE method.  The values given in the parentheses are the standard deviations of the daily predictive standard deviations.  The large day-to-day variation in the sharpness is to be expected as different levels of uncertainty are associated with different prediction scenarios.  In general, the forecasts are more uncertain for higher expected peak wind speeds.  {An alternative approach to assess the sharpness is to calculate the average width of symmetric prediction intervals, see e.g. \cite{Gneiting05b} and \cite{Raftery05}.  Similarly, the average coverage of the prediction intervals may be used to assess calibration.}  

\begin{table}[t]
  \centering
  \caption{Average Brier score at three thresholds and quantile score at four quantiles of peak wind speed predictions at Valkenburg from January 1, 2001 to July 1, 2009.  The GEV methods apply the location parameter in (\ref{eq:optimal location}) and the scale parameter in (\ref{eq:optimal scale}).  The link function for the scale parameter is indicated in parenthesis.  The best performance in each column is indicated in bold. }
  \label{tab:scores}
  \smallskip
  \begin{tabular}{ld{1.2}d{1.2}d{1.2}d{1.2}d{1.2}d{1.2}d{1.2}}
    \toprule 
    & \multicolumn{3}{c}{Brier Score (1/100)} & \multicolumn{4}{c}{Quantile Score (1/10 m/s)} \\
    \cmidrule(r){2-4} 
    \cmidrule(l){5-8}  
    & \multicolumn{1}{c}{14m/s} & \multicolumn{1}{c}{18m/s} & \multicolumn{1}{c}{25m/s}  &   \multicolumn{1}{c}{0.75} & \multicolumn{1}{c}{0.9} & \multicolumn{1}{c}{0.95} & \multicolumn{1}{c}{0.99}\\
    \midrule 
    GEV-MLE (id)       & 6.19 & 3.27 & 0.41 & 4.86 & 2.97 & 1.88 & 0.57 \\
    GEV-CRPS (id)      & 6.18 & \mathbf3.\mathbf2\mathbf1 & \mathbf0.\mathbf3\mathbf9 & \mathbf4.\mathbf8\mathbf1 & \mathbf2.\mathbf9\mathbf5 & \mathbf1.\mathbf8\mathbf7 & \mathbf0.\mathbf5\mathbf6 \\
    GEV-Bayes (log)    & \mathbf6.\mathbf{16} & 3.24 & 0.43 & 4.87 & 2.98 & \mathbf1.\mathbf{87} & 0.58 \\
    lognormal GLM      & 6.55 & 3.70 & 0.66 & 6.73 & 3.47 & 2.20 & 0.69 \\
    \bottomrule  
  \end{tabular}
\end{table}

\begin{table}[t]
  \centering
  \caption{Average skill scores (in percentages) of peak wind speed predictions at Valkenburg from January 1, 2001 to July 1, 2009.  The GEV methods apply the location parameter in (\ref{eq:optimal location}) and the scale parameter in (\ref{eq:optimal scale}).  The link function for the scale parameter is indicated in parenthesis.  The skill scores for the thresholds is based on the Brier score and the skill scores for the quantiles is based on the quantile score.  The reference forecast is the stationary GEV-MLE method.  The best performance in each column is indicated in bold.}
  \label{tab:skill scores}
  \smallskip
  \begin{tabular}{ld{2.1}d{2.1}d{2.1}d{2.1}d{2.1}d{2.1}d{2.1}}
    \toprule 
    & \multicolumn{3}{c}{Threshold} & \multicolumn{4}{c}{Quantile} \\
    \cmidrule(r){2-4} 
    \cmidrule(l){5-8}  
    & \multicolumn{1}{c}{14m/s} & \multicolumn{1}{c}{18m/s} & \multicolumn{1}{c}{25m/s}  &   \multicolumn{1}{c}{0.75} & \multicolumn{1}{c}{0.9} & \multicolumn{1}{c}{0.95} & \multicolumn{1}{c}{0.99}\\
    \midrule 
    GEV-MLE (id)       & 68.8 & 64.0 & 52.5 & 68.7 & 68.2 & 66.7 & 63.7 \\
    GEV-CRPS (id)      & 68.9 & \mathbf6\mathbf4.\mathbf7 & \mathbf5\mathbf4.\mathbf6 & \mathbf6\mathbf9.\mathbf0 & \mathbf6\mathbf8.\mathbf4 & \mathbf6\mathbf7.\mathbf0 & \mathbf6\mathbf4.\mathbf6 \\
    GEV-Bayes (log)    & \mathbf6\mathbf9.\mathbf0 & 64.4 & 50.4 & 68.7 & 68.1 & \mathbf6\mathbf7.\mathbf0 & 63.3 \\
    lognormal GLM      & 67.0 & 59.4 & 23.7 & 56.7 & 62.3 & 61.1 & 56.3 \\
    \bottomrule  
  \end{tabular}
\end{table}

Table~\ref{tab:scores} shows the Brier scores at three high thresholds and the quantile scores for four high quantiles for our four methods. Wind speeds of $14-18$m/s are generally considered as near gale, wind speeds of $18-25$m/s are defined as gale and strong gale, while storm values are $25$m/s and above. The corresponding skill scores are shown in Table~\ref{tab:skill scores} where the reference forecast is the stationary GEV-MLE method.  The standard error for the Brier score is $(0.31,0.25,0.09)$ for the thresholds $(14,18,25)$ and  the standard error for the quantile score is about $0.10$ for each quantile.  These scores measure predictive performance at certain points of the predictive distribution or the quantile function, and may indicate local deficiencies.  The lognormal GLM provides significantly less skill for all thresholds and probabilities, except for the $0.99$ quantile where the difference is no longer significant.  All methods based on a non-stationary GEV distribution perform similarly, with the GEV-CRPS methods showing marginally better performance than the other two methods.

Note that the scores in Table~\ref{tab:scores} cannot be compared across columns as these are based on different subsets of the data.  The apparent improvement in the scores for higher thresholds/quantiles is merely an effect of the decreasing data set on which the values are based.  This is evident if the results of Table~\ref{tab:scores} are compared to the results on Table~\ref{tab:skill scores}.  As the same reference forecast is used for all the columns in Table~\ref{tab:skill scores}, we can compare the results across columns relative to the reference forecast.  Here, we see that the performance of all the methods decreases somewhat with higher thresholds/quantiles compared to the stationary GEV-MLE reference method. 

{Alternatively, weighted proper scoring rules may be applied to focus on areas of special interest in the predictive distribution.  \cite{Diks11} consider weighted versions of the ignorance score while \cite{Gneiting11} discuss approaches to weight the CRPS.  The weight function can here either be based on quantiles using the representation in (\ref{Eq:CRPS quantile}) or it can be based on thresholds using the representation in (\ref{Eq:CRPS}).  Note that the weight functions must be defined independent of the observed value to uphold propriety \citep{Gneiting11}.}     

\subsection{Parameter estimation}
{Verification in terms of scores and residual quantile plots provides average performance of the prediction method.
Another aspect is the variance and covariances of the parameter estimates.
Bayesian prediction provides estimates of the posterior distribution of the parameters, whereas optimum score estimation only gives approximate covariances based on the profile score function (i.e. profile likelihood \citep{Coles01} in maximum likelihood estimation).
An indication of uncertainty in the parameter estimation is provided by the variability of the estimates over the different training data set used in cross-validation.
In the cross-validation procedure we successively remove one year of data, which provides us with nine estimates of each parameter for each method.}

\begin{figure}[t]
\centering
\includegraphics[width=11cm]{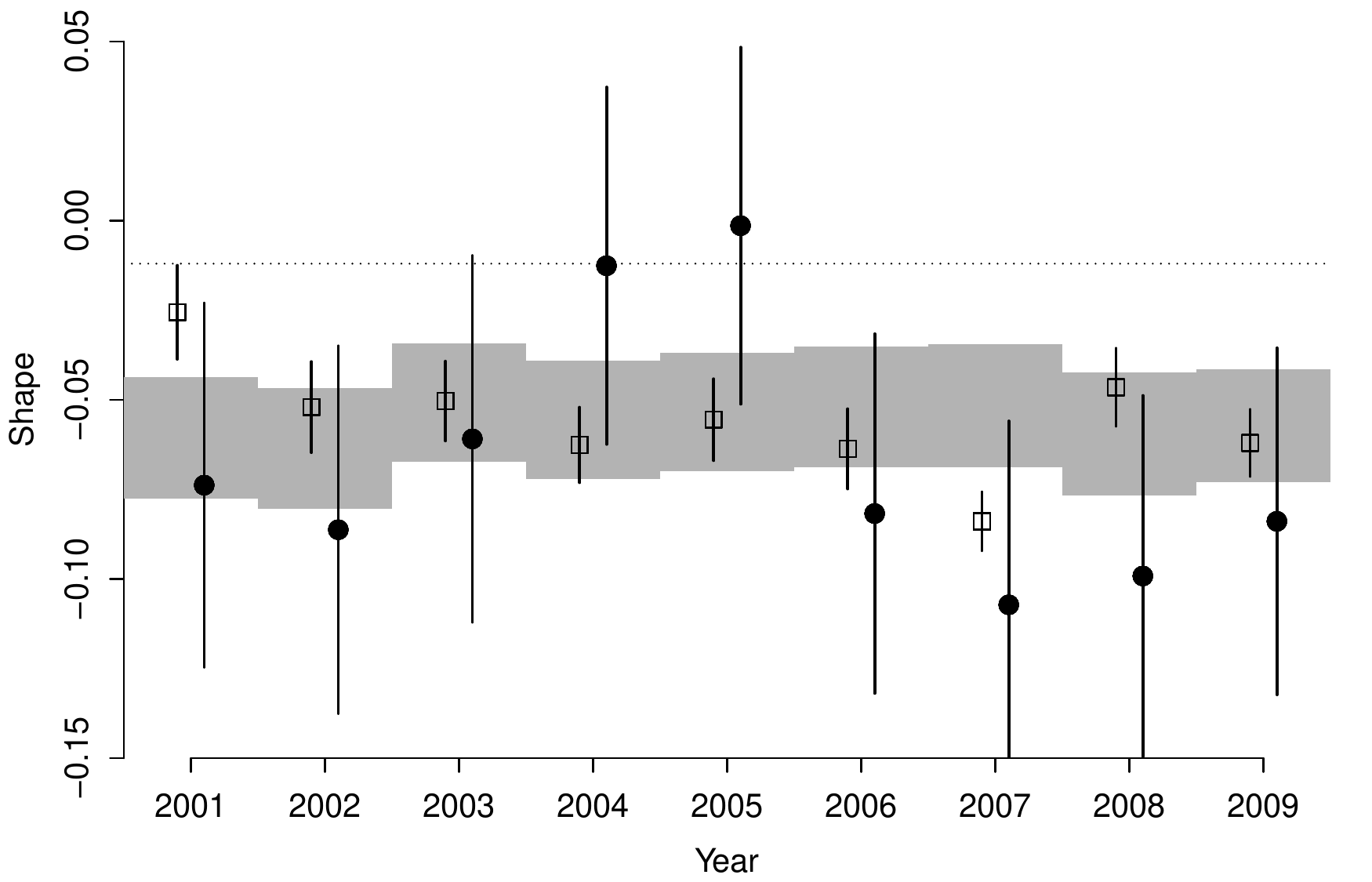}
\caption{The estimated yearly shape parameters for the GEV methods using the location parameter in (\ref{eq:optimal location}) and the scale parameter in (\ref{eq:optimal scale}).  The gray boxes indicate the $90\%$ posterior intervals for for $\xi$ under the Bayesian model, the black dots ($\bullet$) indicate the parameter estimates under the GEV-CRPS(id) model, and the squares ($\Box$) indicate the parameter estimates under the GEV-MLE(id) model.  The vertical lines indicate the respective standard errors.  For comparison, the shape parameter of the observations estimated via MLE, $\hat{\xi} = -0.013$, is indicated with a dotted line.}
\label{Fig:shape}
\end{figure}

Figure~\ref{Fig:shape} shows the parameter estimates for the yearly shape parameter $\xi$ under the three GEV methods discussed in the previous section.  The $90\%$ posterior intervals obtained with the Bayesian method are fairly constant between years, while there is a large variation in the estimated values under both GEV-MLE and GEV-CRPS.  Furthermore, the standard errors are significantly greater for the GEV-CRPS method. This is in accordance with Fig. \ref{Fig:CRPS} where the minimum of the CRPS is less sensitive to changes in $\xi$ and the discussion related to Fig. \ref{Fig:CRPSGEV}. The MLE and CRPS optimum score estimates of $\xi$ are uncorrelated which suggests that the variability in the estimates is not due to interannual changes in the data. The majority of the estimates are significantly negative, indicating that the non-stationary GEV models are of Weibull type, in contrast to  the goodness of fit analysis performed in Section~\ref{Sec:Data}, where $\hat \xi$ is only slightly negative. 

\begin{figure}[t]
\centering
\includegraphics[width=0.99\textwidth]{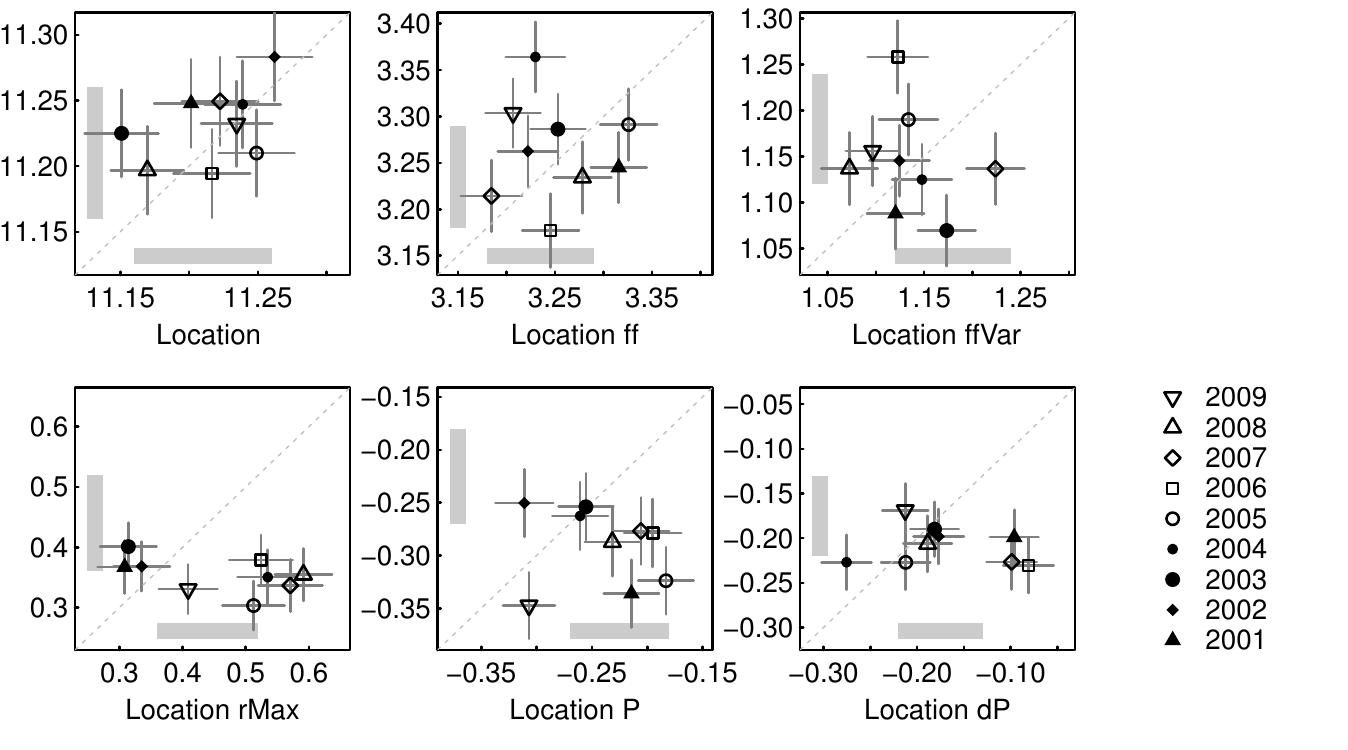}
\caption{Optimum score estimates of yearly location parameters for the GEV methods using the location parameter in (\ref{eq:optimal location}) and the scale parameter in (\ref{eq:optimal scale}).  The dots represent the maximum likelihood estimates (x-axis) and the optimum CRPS estimates (y-axis) for the different years. The gray lines indicate the respective standard errors.  The gray bars on each axis indicate the $90\%$ posterior intervals for the parameters under the Bayesian model.}
\label{Fig:location}
\end{figure}

The parameter estimates for the location coefficients in (\ref{eq:optimal location}) are shown in Figure~\ref{Fig:location}.  For clarity of the presentation and as the $90\%$ posterior intervals are very similar for each run of the Bayesian method, we only show the $90\%$ posterior interval of the joint posterior sample from all years.  In general, there is no apparent correlation between the maximum likelihood estimates and the optimum CRPS estimates.  Furthermore, the estimates for both methods are quite variable for different training sets and this variability cannot be explained solely by the uncertainty in the estimation which is of similar magnitude for both methods and across training sets.  All three methods display the largest uncertainty in the estimate for the intercept coefficient. All methods show a strong positive dependence of the GEV location parameter on $\ff$ and $\ffVar$, a weaker dependence on $\rMax$, and a negative dependence on pressure $p$ and pressure tendency $\dP$.

The Bayesian posterior distributions for the scale coefficients in (\ref{eq:optimal scale}) cannot be directly compared to the corresponding estimates under the frequentist approaches as we have applied the identity link for the frequentist approaches while we use a logarithmic link for the Bayesian method.  As for the location coefficients, no apparent correlation can be found between the MLE coefficient estimates and the minimum CRPS estimates and the variability between training sets is generally greater than expected based on the estimation uncertainty alone. Like the location parameter, the non-stationary GEV scale parameter positively depends on $\ff$, $\ffVar$, and $\rMax$. Hence, predictive uncertainty increases for large expected peak wind speed.

\section{Discussion}\label{Sec:Discussion}

As e.g. \citet{Dawid84} and \citet{Gneiting08} have argued before us, predictions for uncertain events ought to include an estimate of the associated uncertainty.  This is easily obtained within a probabilistic framework where the predictions take the form of probability distributions.  Proper scoring rules such as the Brier score or the CRPS are widely used to assess the predictive skill in probabilistic weather forecasting, see e.g. \citet{Wilks11}.  In this paper, we discuss how proper scoring rules may be used for out-of-sample model selection and verification for extreme value distributions.  We present a closed-form expressions of the CRPS for this class of distributions.  With these expressions at hand, the CRPS may be used for optimum score estimation as an alternative to maximum likelihood estimation or Bayesian inference.  

In a case study, we compare various approaches to derive non-stationary distributions for daily peak wind speed observations.  The non-stationarity is built in by including covariates, i.e. meteorological parameters that are observed together with the gust measurement.  Our competing methodologies comprise a lognormal GLM for the wind gust factor and non-stationary GEV models for the daily wind gusts.  The {non-stationary} GEV approaches provide significantly higher skill than the {corresponding} lognormal GLM.  This confirms findings in \citet{Friederichs09} that extreme value theory provides an appropriate and theoretically consistent statistical model for wind gusts.  

Due to the close relationship between wind speed and peak wind, mean wind speed is the main covariate in our model.  However, the predictive performance is significantly improved by also including information on additional weather variables.  The most informative covariates besides mean wind speed are the variability of the mean wind speed, the maximum rain rate, and the pressure tendency throughout the day.  With six potential covariates for both the location and the scale parameter, our model space includes a total of $2^{12}$ models.  This large space can easily be searched using a Bayesian variable selection procedure which returns different covariate sets for the location and for the scale.  Overall, the robustness and the predictive performance of the Bayesian framework is very good though it should be noted that the Bayesian inference is considerably slower than the frequentist methods. 

\cite{Gneiting05b} compare minimum CRPS estimation and maximum likelihood estimation in a prediction framework and show that the former returns forecasts with slightly better overall predictive performance and significantly improved calibration.  Their results confirm well with our findings in that the optimum CRPS estimation outperforms the maximum likelihood approach in nearly all aspects of our verification even though the differences are sometimes small.  A drawback of the minimum CRPS estimation is its weak discriminant power with respect to the GEV shape parameter. 

{In our case study, we have performed an extensive analysis of the properties of the various scoring rules and associated inference methods for the GEV distribution. The GEV is the self-evident distribution in the case of block maxima, i.e. in the case of peak wind speed. In the case of threshold excesses, a peak-over-threshold (POT) or Poisson point process approach would be more appropriate.  We assume that similar results will hold for the POT approach using a GPD distribution and the closed-form expression given in this paper. Since the Poisson point process is generally represented in terms of GEV parameters, optimum score estimation using the CRPS should be possible as well.}

\section*{Acknowledgments} 
Special thanks go to Michael Scheuerer for valuable discussions and comments on early versions of the manuscript. The authors also thank Tilmann Gneiting, Andreas Hense, Alex Lenkoski, and Michael Weniger for helpful discussions, and the associate editor and an anonymous reviewer for their useful comments. We acknowledge the support of the Volkswagen Foundation through the project ``Mesoscale Weather Extremes - Theory, Spatial Modeling and Prediction (WEX-MOP)''.

\clearpage
\bibliography{artikel}

\appendix

\clearpage
\section{Derivation of the CRPS for a GEV}\label{A:CRPS_GEV}
For the deviation of the CRPS under the GEV, we apply the representation in (\ref{Eq:CRPS quantile}), where the CRPS is presented as an integral over the quantile score in (\ref{Eq:quantile score}).  We can write the quantile score as 
\[
S_Q^\tau(F,y) = \tau \big[ y - F^{-1}(\tau)\big] - \mathbbm{1}\{ \tau \geq F(y)\} \big[ y - F^{-1}(\tau)\big]
\]
from which it is easily seen that 
\begin{equation}\label{Eq:CRPS2}
S_{CRP}(F,y) = y \big[ 2F(y) -1 \big] -2  \int_{0}^1 \tau F^{-1}(\tau) d \tau + 2 \int_{F(y)}^1 F^{-1}(\tau) d \tau. 
\end{equation}
The quantile function of the GEV is given by 

\begin{equation}\label{Eq:InvGEV}
    F_{GEV}^{-1}(\tau) =  \left\{ \begin{array}{lc}
      \mu - \frac{\sigma}{\xi} \big[ 1- \left( -\log \tau 
\right)^{-\xi} \big] ,&  \xi \ne 0 \\[0.2cm]
     \mu - \sigma \log \left( -\log \tau  \right),& \xi = 0  \end{array}
\right. .
\end{equation}
For a non-zero shape parameter $\xi \ne 0$ and with $F = F_{GEV_{\xi \ne 0}}$, we obtain
\begin{equation}\label{eq:int1 nonzero}
2 \int_{0}^1 \tau F^{-1}(\tau) d \tau = \mu - \frac{\sigma}{\xi} + 2 \int_0^1 \tau \big( - \log \tau\big)^{-\xi} d \tau
\end{equation}
and 
\begin{equation}\label{eq:int2 nonzero}
2 \int_{F(y)}^1 F^{-1}(\tau) d \tau = 2 \Big[ \mu - \frac{\sigma}{\xi} \Big] \big[ 1 - F(y) \big] + 2 \frac{\sigma}{\xi} \int_{F(y)}^1 \big( - \log \tau\big)^{-\xi} d \tau.   
\end{equation}
To solve the two remaining integrals in (\ref{eq:int1 nonzero}) and (\ref{eq:int2 nonzero}), we use that 
\[
\int \tau (-\log \tau)^{-\xi} d\tau = 2^{\xi-1} \Gamma_u(1-\xi,-2\log \tau)
\]
and 
\[
\int (-\log \tau)^{-\xi} d\tau = \Gamma_u(1-\xi,-\log \tau),
\] 
where $\Gamma_u(a, \tau)=  \int_{\tau}^\infty t^{a-1} e^{-t} dt$ is the upper incomplete gamma function.  By combining (\ref{Eq:CRPS2}), (\ref{eq:int1 nonzero}), (\ref{eq:int2 nonzero}), and the integral equations above, we get
\[
S_{CRP}(F_{GEV_{\xi \neq 0}},y)
        = \Big[ y - \mu + \frac{\sigma}{\xi} \Big] \big[ 2F_{GEV_{\xi \ne 0}}(y)-1 \big]
         - \frac{\sigma}{\xi} \Big[ 2^{\xi}   \Gamma(1-\xi)
                - 2\Gamma_l\big(1-\xi, -\log F_{GEV_{\xi \ne 0}}(y) \big) \Big],
\]
where $\Gamma_l(a, \tau) = \Gamma(a)-\Gamma_u(a,\tau)$ is the lower incomplete gamma function.

For a Gumbel type GEV with $\xi=0$, similar calculations show that 
\begin{align}\label{eq:CRPS3}
S_{CRP}(F_{GEV_{\xi = 0}},y)
=  & [y - \mu] \big[2F_{GEV_{\xi = 0}}(y) - 1 \big] + 2\sigma \int_0^1 \tau
\log (-\log \tau) d \tau \\
& - 2\sigma \int_{F_{GEV_{\xi = 0}}}^1 \log(-\log \tau) d \tau. \nonumber
\end{align}
The first integral is solved as
\begin{equation*}
 \int \tau \log(-\log \tau) d\tau  = \frac{1}{2}\Big[\tau^2 \log(-\log \tau) - Ei (2
\log \tau) \Big] 
\end{equation*}
where $Ei(x)=  \int_{-\infty}^x \frac{e^t}{t} dt$ is the exponential
integral also given by 
\begin{equation*}
  Ei(x ) = C + \log(|x|) + \sum_{k=1}^\infty
  \frac{x^k}{k! \, k}
\end{equation*}
with $C \approx 0.5772$ being the Euler-Mascheroni constant. Further, we use 
\begin{equation*}
    \int \log(-\log \tau ) d \tau = \tau \log(-\log \tau ) - Ei(\log \tau) ,
\end{equation*}
to solve the second integral in (\ref{eq:CRPS3}).  This leads to a closed-form expression for the CRPS under a Gumbel-type GEV with
\begin{align*}
  S_{CRP}(& F_{GEV_{\xi = 0}},y) \\
  & =  [y - \mu] \big[2F_{GEV_{\xi = 0}}(y) - 1 \big] \\
  & \quad + \sigma \lim_{\nu \rightarrow 1} \Big[ \nu^2 \log(-\log \nu ) - Ei(2 \log \nu ) \Big] - \lim_{\eta \rightarrow 0}\Big[ \eta^2 \log(-\log \eta ) - Ei(2 \log \eta ) \Big] \\
  & \quad - 2\sigma \lim_{\nu \rightarrow 1} \Big[ \nu
  \log ( -\log \nu ) - Ei(\log \nu ) \Big] - 2 \sigma  
  \Big[ F_{GEV_{\xi=0}}(y)  \frac{y - \mu}{\sigma} + Ei\big(\log F_{GEV_{\xi=0}}(y) \big) \Big] \\
  & = - (y - \mu) - 2 \sigma Ei\big(\log F_{GEV_{\xi=0}}(y) \big) \\
  & \quad + \sigma \lim_{\nu \rightarrow 1} \Big[ (\nu-1)^2 \log (-\log \nu) + C - \log 2 + \sum_{k=1}^{\infty} \frac{(2 \log \nu)^k + 2 (\log \nu)^k}{k! k}  \Big]\\
  &=  -\ (y - \mu) - 2 \sigma Ei \big(\log F_{GEV_{\xi = 0}}(y)\big) + \sigma[C-\log2],
\end{align*}
where we use that $\lim_{\eta \rightarrow - \infty}Ei(\eta) = 0$.  The series expansion to compute $Ei ( \log F_{GEV_{\xi = 0}}(y))$ converges rapidly for $F_{GEV_{\xi = 0}}(y)$ much greater than $0$.  However, it may fail for very small values of $F_{GEV_{\xi = 0}}(y)$.  We use the GNU Scientific Library \citep{Galassi10} to compute $Ei(\cdot)$. 

\section{Derivation of the CRPS for a GPD}\label{A:CRPS_GPD}

The derivation of the closed-form expression of the CRPS for a GPD is very similar to the derivations for a GEV in Appendix A with significantly simpler calculations.  The GPD quantile function is given by 
\[
F_{GPD}^{-1}(p) = \left\{ \begin{array}{lc}
    u- \frac{\sigma_u}{\xi} + \frac{\sigma_u}{\xi}(1-p)^{-\xi} ,&  \xi \ne 0 \\
    u- \sigma_u \log(1-p) ,& \xi = 0  \end{array} \right. .
\]
By applying the representation of the CRPS given in (\ref{Eq:CRPS2}), we get
\begin{align*}
  S_{CRP}(F_{GPD_{\xi \neq 0}},y)  
  &=   \Big[ y - u+\frac{\sigma_u}{\xi} \Big] \big[ 2F_{GPD_{\xi \neq 0}}(y) - 1 \big] \\
  & \quad - \frac{2 \sigma_u}{\xi} \Big[ \int_{F_{GPD_{\xi \neq 0}}(y)}^1 (1-\tau)^{-\xi} d \tau - \int_0^1 \tau (1-\tau)^{-\xi} d \tau \Big] \\ 
  &=   \Big[ y - u+\frac{\sigma_u}{\xi} \Big] \big[ 2F_{GPD_{\xi \neq 0}}(y) - 1 \big] \\
  & \quad - 2 \frac{\sigma_u}{\xi (\xi-1)} \Big[ \frac{1}{\xi-2} + \Big[ 1-F_{GPD_{\xi \neq 0}}(y) \Big] \Big[ 1+\xi\frac{y-u}{\sigma_u} \Big] \Big].
\end{align*}
Here, the second integral can be solved using integration by parts.  For $\xi = 0$, we obtain the following expression
\begin{align*}
  S_{CRP}(F_{GPD_{\xi = 0}},y)  
  &=   (y-u)\big[ 2F_{GPD_{\xi = 0}}(y) - 1 \big] \\
  & \quad + 2 \sigma_u \Big[ \int_0^1 \tau \log(1-\tau) d \tau - \int_{F_{GPD_{\xi = 0}}(y)}^1 \log(1-\tau) d \tau \Big] \\
  &= y-u - \sigma_u\Big[ 2F_{GPD_{\xi = 0}}(y) -\frac{1}{2} \Big], 
\end{align*}
where we use the integral equation
\[
 \int \log(1-x) dx = x \big[ \log(1-x) - 1 \big] - \log(1-x)
\]
and integration by parts. 

\end{spacing}

\end{document}